\documentclass[twoside,fleqn]{article}
\usepackage{psfig}
\usepackage{epsfig}
\usepackage{espcrc2}

\newcommand{\be}{\begin{equation}}
\newcommand{\ee}{\end{equation}}

 1
 1
 1

\title{Spectral Features and Masses in the PeV Region}
\author{A.D.Erlykin\address{P.N.Lebedev Physical Institute - Moscow, Russia} 
\thanks{Corresponding author. Tel +7 095 1358737, fax +7 95 1357880, e-mail erlykin@sci.lebedev.ru},
A.W. Wolfendale\address{Department of Physics, University of
Durham - Durham, UK}}

\begin{document}

\begin{abstract}
An analysis is made of the masses and spectral features for cosmic
rays in the PeV region, insofar as they have a bearing on the problem
of the interaction of cosmic ray particles.

In our Single Source Model we identified two 'peaks' seen in a summary
of the world's data on primary spectra, and claimed that they are
probably due to oxygen and iron nuclei from a local, recent
supernova. In the present work we examine other possible mass
assignments. We conclude that of the other possibilities only Helium
and Oxygen ( instead of O and Fe ) has much chance of success; the
original suggestion is still preferred, however. Concerning our
location with respect to the SNR shell, the analysis suggests that we
are close to it - probably just inside.     
\end{abstract}
\maketitle

\section{Introduction}

In our Single Source Model (~updated version is in \cite{EW1}~)
we explained the knee as the effect of a local, recent supernova, the
remnant from which accelerated mainly oxygen and iron. These nuclei
form the intensity 'peaks' which perturb the total 'background'
intensity. The comprehensive analysis of the world's data gives as our
datum the plots given in the Figure 1; these are 'deviations from the
running mean' for both the energy spectrum mostly from Cherenkov data
and the summarised electron size spectrum. It is against these datum
plots that our comparison will be made. 

In the present work we endeavour to push the subject forward by
examining a number of aspects. They are examined, as follows: \\
(i) Can we decide whether the solar system is inside the supernova
shock or outside it ? \\
(ii) Is the identification of Oxygen and Iron in the peaks correct ?
\\
(iii) Can both the peaks be due to protons rather than nuclei ? In
view of claims from a few experiments (~DICE, BLANCA~) that the mean
mass is low in the PeV region, it is wise to examine this
possibility. 
\section{The Solar System's position with respect to the nearby SNR}
The appreciation that the frequency of SN in the local region of the
Interstellar Medium (~ISM~) has been higher than the Galactic average,
over the past million years, has improved the prospects for the SSM
being valid \cite{EW2,EW3} and thereby increases the probability that
we are close to the surface of a remnant.

It is doubtlessly possible for particles to escape from an SNR shock
and propagate ahead. Such a situation has been considered in the 
'Berezhko-model'. The problem concerns uncertainties in the diffusion
coefficient for the ISM; however, estimates have been made
\cite{Bere1,Bere2} and Figure 1 shows the result for the Sun being
outside the shock at the distance of 1.5$R_S$ for the center of SNR
(~$R_S$ is the radius of the remnant~).
\begin{figure}[ht]
\psfig{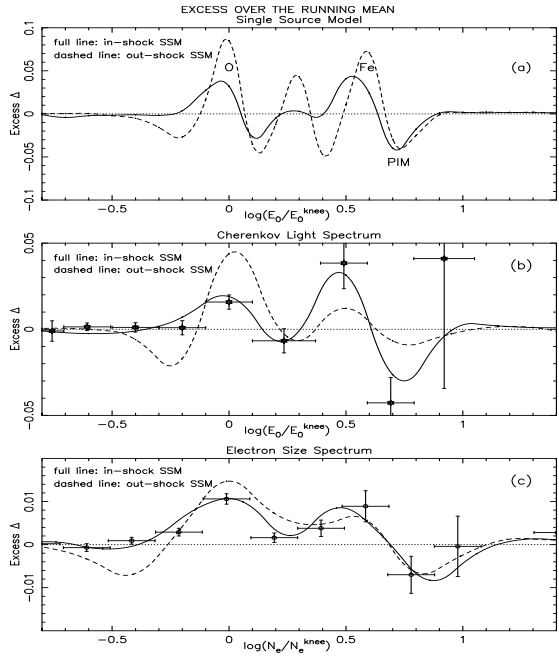}
\caption{\footnotesize Excess over the running mean for (a) the primary SSM
energy spectrum, (b) averaged Cherenkov light spectrum; (c) averaged
EAS electron size spectrum. The results relate to the average excess
in $\Delta log(E/E^{knee})$ = 0.2 bins for (b) and (c), for (a) the bin
size is 0.1. The full curve is for the case where the Sun is inside
the shock as in the original SSM. The dashed line is for the Sun
outside of the shock. Curves for SSM and Cherenkov light are without
noise corrections, whereas for the electron size spectrum a noise
correction, using $\sigma(logN_e)$ = 0.16 has been made. }
\label{fig1}
\end{figure}

It is seen that the result does not fit well the datum points at all. 
The model tested must be rejected in its given form. 

It is possible to restore it by taking an energy spectrum of more nearly the
form for the 'inside SNR' location or at the position outside, but
very close to the shell. The corresponding cureves are shown in Figure
1 by full lines.
\section{SNR: Helium and Oxygen}
A tolerable astrophysical case could be made for helium and oxygen
rather than oxygen and iron, and the direct measurements at lower
energies than the knee region do not really rule it out.

Figure 2 shows the $\Delta$-values for the corresponding spectra. The
separation of the He and O peaks is a little greater than for O and Fe
(~8/2 compared with 26/8~) and this causes the He, O pattern to be
displaced somewhat. Although the fit to the datum points is not as
good as for O, Fe, the He, O combination cannot be ruled out on the
basis of the $\Delta$-plots alone. The absence of the preferred-by-us
nuclei between the two peaks is a worry, though (~incertion of carbon
does not help to fill the gap between two peaks~). The Fe peak would
then be expected at log($N_e/N_e^{knee}$) = 1.1.
\begin{figure}[ht]
\psfig{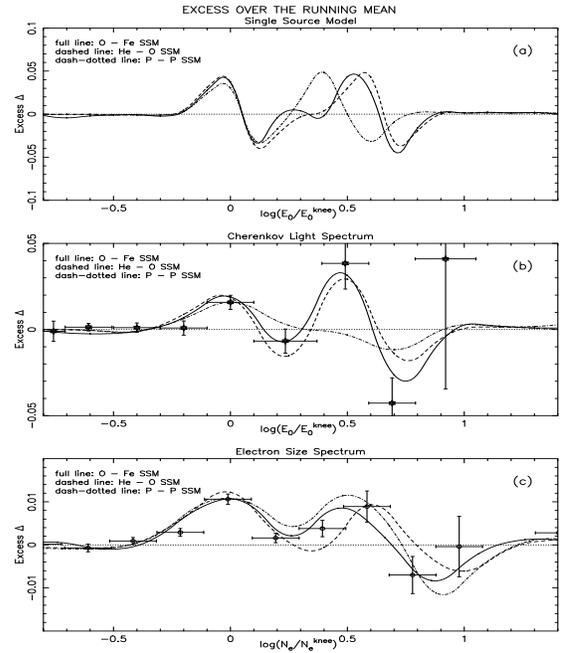}
\caption{\footnotesize Excess over the running mean for the usual
assumption: O, Fe (~full line~), for He, O (~dashed line~) and for P-P
(~dash-dotted line~). Although O, Fe gives the best fit, He, O cannot
be ruled out, P - P is disallowed.}
\label{fig2}
\end{figure}
\section{Proton Peaks}
Calculations have been made for the case of two proton peaks, the
proton spectra having been taken to be the standard interior-to-the SNR
form. The result is also shown in Figure 2.

An interesting situation develops here. Although it is possible to
tune either the energy spectrum or the size spectrum to fit the 
$\Delta$-results, it is not possible to choose an energy spectrum
which fits both. This arises because of the sensitivity of the number
of electrons at the detection level to the primary mass. In Figure 2
the separation of the proton peaks in the energy spectrum was chosen
such that the $\Delta$-distribution for shower size was a reasonable
fit to the data. However, the separation of the peaks in the energy
spectrum necessary for the shower size fit is less than that for O,Fe
by 0.15; the result is that after the necessary binning (~0.2 in
$logE$ units~) for the energy spectrum there is no agreement there.
\section{Discussion about the Nature of the Peaks and our Location}
It is evident from the foregoing that the two-proton peak model is 
unacceptable. This result cast doubt on the analyses of EAS data which
conclude that the mean primary mass is 'low' 
(~$\langle lnA \rangle \simeq 1.5$~) in the PeV region. As mentioned
already, it is our view that some, at least, of the models used in the
mass analyses are inappropriate for the interactions of nuclei,
particularly for the production and longitudinal development of the
electromagnetic component. It is interesting to know, in connection
with mean mass estimates, that the recent work using the Tibet EAS
array \cite{Ameno} has given strong support for the result - favoured
by us - in which the average cosmic ray mass increases with energy. In
fact, their mass is even higher than ours: $\langle lnA \rangle \simeq
3.1$, compared with our 2.4, at 1 PeV, and 3.3, compared with 3.0 at
10 PeV. Equally significant is the fact that the sharpness of the iron
component that they need to fit the overall data is quite
considerable: $S$ = 1.4. It will be remembered that straightforward 
Galactic diffusion - the 'conventional model' - gives $S \simeq 0.6$
for any one mass component and $S \simeq 0.3$ for the whole spectrum 
\cite{EW4}.

Returning to the question of 'our' location with respect to the SNR it
seems difficult to account for the $\Delta$-distribution if we are
some distance outside the shell, unless the diffusion coefficient for
cosmic ray propagation in the ISM is almost energy-independent. We
appear to be inside, or only just outside.

Finally, concerning the nature of the peaks: O, Fe or He, O, it is
difficult to rule out the latter from the $\Delta$-plots alone,
although the lack of an iron peak is surprising. However, there is
some evidence from the Tunka-25 Cherenkov experiment for a further
peak at roughly the correct energy for the third (~Fe~) peak
\cite{Budne}. There is also a hint of a peak in KASCADE
spectrum, which is observed at an even higher energy than in Tunka-25 
\cite{Schat}. Most other experiments - but not all - do not have the
sensitivity to detect a further peak so the situation here is still
open.

We still prefer our original suggestion, viz. that the peaks are due
to O and Fe, and their shape is the consequence of the sharp cut-off
in the energy spectrum of particles accelerated by SNR. The main
reason for the preference is the fact that O and Fe spectra
extrapolate and fit direct measurements of those components rather
well \cite{EW5} and there are good astrophysical reasons favouring
these nuclei.
\section{Conclusions}
The Single Source Model, with its explanation of the knee in the
cosmic ray energy spectrum in terms of particles (~probably
principally nuclei of oxygen and iron~) from a recent, local SN, has
been examined further. It is true that the identity of the nuclei is
not completely secure and it is just possible that rather than O, Fe,
the combination is He, O: however, we still prefer the original
explanation.

The question of the nature of the particles responsible for the knee
is, therefore, still somewhat uncertain; however, that there is
structure in the spectrum, indicative of a single source, seems to be
rather secure.

Turning to our location, the analysis suggests that we are just inside
the shell, although, with a different diffusive mode of propagation
for the particles we could be just outside it.

\vspace{0.5cm}

{\bf Acknowledgements}

\vspace{0.2cm}

The Royal Society and the Particle Physics and Astronomy Research
Council are thanked for their financial support.

\end{document}